# Barrier to recombination of oppositely charged large polarons

David Emin

Department of Physics and Astronomy

University of New Mexico

Albuquerque, New Mexico, 87131, USA

Abstract

Electronic charge carriers in ionic materials can self-trap to form large polarons. Interference between the ionic displacements associated with oppositely charged large polarons increases as they approach one another. Initially this interference produces an attractive potential that fosters their merger. However, for small enough separations this interference generates a repulsive interaction between oppositely charged large polarons. In suitable circumstances this repulsion can overwhelm their direct Coulomb attraction. Then the resulting net repulsion between oppositely charged large polarons constitutes a potential barrier which impedes their recombination.

1. Introduction

Perovskite materials are ionic solids with especially displaceable ions as evidenced by the very large ratios of the static to high-frequency dielectric constants in these solids.[1-4] The long-range electron-phonon interactions between displaceable ions and electronic charge carriers enables their forming large polarons.[5,6] Moderate large-polaron mobilities result from the extremely weak acoustic-phonon scattering of these slow moving heavy-massed quasi-particles.[7-9] Attractive interactions between like-charged large polarons facilitate their pairing to form large bipolarons.[10,11] While the Pauli exclusion principle precludes bipolarons merger into even grander polarons, the phonon-mediated attraction between large bipolarons facilitates their condensation into a liquid.[12,13] Novel superconductivity can result from the Bose condensation of a large-bipolaron liquid.[14]

Here it is shown that polaron effects also generate a repulsive interaction between oppositely charged large polarons. With a sufficiently large static dielectric constant [e.g. Ref. (4) reports $\varepsilon_0 \approx 1000$ in $MAPbX_3$] this repulsion can dominate the direct Coulomb attraction between oppositely charged carriers. Such a net repulsion may explain the exceptionally weak recombination observed in perovskite solar cells that underlies their exceptionally high energy-conversion efficiencies.[15] Indeed, the hybrid organic-inorganic lead halide perovskites used in solar cells are reported to have (1) huge ratios of their static to high-frequency dielectric constants and (2) electronic charge carriers that form large polarons.[4,16,17]

2. Calculation

To address recombination of oppositely charged large polarons, the net polaron contribution to their adiabatic energy as a function of their separation $s$ is computed. Following the classical version of the Emin-Holstein procedure, where electronic carriers' wavefunctions are replaced by Kronecker delta functions in position, the carriers' potential energy and the material's strain energy are written as

$$V(r_1, r_2) = \int du\, [Z(r_1 - u) + Z(r_2 - u)]\Delta(u) \quad (1)$$

and

$$S = \frac{k}{2V_c} \int du\, \Delta^2(u), \quad (2)$$

where $Z(r - u)$ denotes the electron-phonon interaction.[5] This function describes the dependence of the potential energy of a carrier at position $r$ on a deformation of the material of $\Delta(u)$ at position $u$. The Hooke's law stiffness constant is given by $k$ and $V_c$ represents the volume of a unit cell.

Minimizing the sum of these two deformation-related contributions to the potential energy yields an expression for the minimum-energy carrier-induced deformation:

$$\Delta_{min}(u) = -\frac{Z(r_1 - u) + Z(r_2 - u)}{k/V_c}. \quad (3)$$

The corresponding minimum of the net potential energy is:

$$V_{ad}(r_1, r_2) = -\frac{V_c}{2k} \int du\, [Z(r_1 - u) + Z(r_2 - u)]^2. \quad (4)$$

The Fröhlich long-range electron-phonon interaction of an electric dipole centered at position $u$ with the electric field generated by a charge carrier at position $r$ is

$$Z_{LR}(r - u) \equiv \frac{\sqrt{\frac{e^2}{4\pi}\left(\frac{1}{\varepsilon_\infty} - \frac{1}{\varepsilon_0}\right)\frac{k}{V_c}}}{|r - u|^2} \cos(\theta), \quad (5)$$

where $\theta$ denotes the angle being the electric dipole and the position vector $r - u$ with $\varepsilon_\infty$ and $\varepsilon_0$ respectively representing the material's high-frequency and static dielectric constants.[6] For oppositely charged carriers that both interact with the material via this interaction the net polaron contribution to the potential energy becomes

$$V_{ad}(r_1, r_2) = -\frac{1}{8\pi}\left(\frac{1}{\varepsilon_\infty} - \frac{1}{\varepsilon_0}\right) \int du\, |E(u; r_1, r_2)|^2, \quad (6)$$

where $E(u;r_1,r_2)$ represents the electric field at $u$ that would be generated in free space by large polarons centered at $r_1$ and $r_2$ having charge $e$ and $-e$. The spatial extents of the large polarons' self-trapped electron and hole, characterized by radii $R_-$ and $R_+$, are inversely proportional to these electronic carriers' effective masses. The solid's electric dipoles are presumed to align with the electric field generated by the self-trapped electronic carriers outside of the regions they occupy. Concomitantly, the integration excludes regions within these large polarons' radii where cancellations diminish electric-dipoles' alignments.

For simplicity, the electric field is taken to be that generated by two oppositely charged points located at $r_1$ and $r_2$. In particular, cylindrical coordinates are employed and the positively charged and negatively charged points are located at $z = S/2$ and $z = -S/2$, respectively. The electric field they generate has components in the radial and $z$ directions that are respectively:

$$E_r(S;r,z)/e = r\left\{\left[r^2 + \left(z - \frac{S}{2}\right)^2\right]^{-3/2} - \left[r^2 + \left(z + \frac{S}{2}\right)^2\right]^{-3/2}\right\} \quad (7)$$

and

$$E_z(S;r,z)/e = \left(z - \frac{S}{2}\right)\left[r^2 + \left(z - \frac{S}{2}\right)^2\right]^{-3/2} - \left(z + \frac{S}{2}\right)\left[r^2 + \left(z + \frac{S}{2}\right)^2\right]^{-3/2}. \quad (8)$$

The radial components of the electric field generated by the two point charges always oppose one another. By contrast, the $z$-components of their electric fields reinforce one another between the two point charges, $S/2 > z > -S/2$. Squaring and summing these two quantities yields:

$$E^2(S;r,z)/e^2 = \left[r^2 + \left(z - \frac{S}{2}\right)^2\right]^{-2} + \left[r^2 + \left(z + \frac{S}{2}\right)^2\right]^{-2} - 2\frac{r^2 + z^2 - \left(\frac{S}{2}\right)^2}{\left\{r^4 + 2r^2\left[z^2 + \left(\frac{S}{2}\right)^2\right] + \left[z^2 - \left(\frac{S}{2}\right)^2\right]^2\right\}^{3/2}}. \quad (9)$$

This function has a saddle-point at the origin with the square of the magnitude of the electric field falling off radially while rising in the positive and negative $z$-directions toward the two point charges: $E^2(S;r,z)/e^2 \approx (64/S^4)[1 - 12(r^2 - 2z^2)]$.

Attention is now focused on the regime depicted in Fig. 1 in which the separation between the two large polarons exceeds the sum of their radii: $S > R_+ + R_-$. As described by Eq. (6), evaluation of $V_{ad}(r_1, r_2)$ proceeds by integrating $E^2(S;r,z)$ of Eq. (9) over all space with the exclusion of spheres of radii $R_+$ and $R_-$ centered at $z = S/2$ and $z = -S/2$, respectively. This integration is performed by dividing the $z$-direction into five regions: 1) above the upper sphere, (2) outside of the upper sphere, (3) between the upper and lower spheres, (4) outside of the lower sphere, and (5) below the lower sphere. The five corresponding domains are defined by: (1) $\infty > z \geq S/2 + R_+$ with $\infty > r \geq 0$; (2) $S/2 + R_+ \geq z \geq S/2 - R_+$ with $\infty > r \geq \sqrt{R_+^2 - (z - S/2)^2}$; (3) $S/2 - R_+ \geq z \geq -S/2 + R_-$ with $\infty > r \geq 0$; (4) $-S/2 + R_- \geq z \geq -S/2 - R_-$ with $\infty > r \geq \sqrt{R_-^2 - (z + S/2)^2}$; (5) $-S/2 - R_- \geq z \geq -\infty$ with $\infty > r \geq 0$.

Adding the five integrations performed for each of the three contributions to $E^2(S;r,z)$ of Eq. (9) yields 15 terms. Many of these terms can be combined. The result is that the adiabatic potential for a positively charged large polaron of radius $R_+$ centered at $S/2$ with a negatively charged large polaron of radius $R_-$ centered at $-S/2$ is:

$$V_{ad}(R_+, R_-, S) \equiv V(R_+) f\left(\frac{S}{R_+}\right) + V(R_-) f\left(\frac{S}{R_-}\right). \quad (10)$$

Here the potential energy of an isolated large polaron of radius $R$ is denoted by

$$V(R) \equiv -\frac{e^2}{2}\left(\frac{1}{\varepsilon_\infty} - \frac{1}{\varepsilon_0}\right)\frac{1}{R}. \quad (11)$$

In addition, the dependence of the adiabatic potential energy of the two oppositely charged large polarons on the separation between them is governed by the interaction function:

$$f(s) \equiv 1 - \frac{1}{2(s^2-1)} + \frac{1}{8s} \ln\left[\frac{(s+1)^2}{(s-1)^2}\right] + \frac{4}{s^2-4}$$
$$+ \int_{-1}^{1} du \left[\frac{1}{(2u+s)^2}\right]\left[\frac{1-2u^2-su}{(1+2su+s^2)^{1/2}} - 1\right]. \quad (12)$$

Pairs of oppositely signed singularities occur for the second and third terms at $s = 1$ and for the fourth and fifth contributions at $s = 2$. These artificial features result from discontinuities at the boundaries of the five regions of integration. These artificialities occur away from the regime in which $f(s)$ will be employed, $s \gg 1$.

In the limit of $s \to \infty$, the potential energy is just that of independent positively charged and negatively charged large polarons:

$$\lim_{s \to \infty} V_{ad}(R_+, R_-, S) = -\frac{e^2}{2}\left(\frac{1}{\varepsilon_\infty} - \frac{1}{\varepsilon_0}\right)\left(\frac{1}{R_+} + \frac{1}{R_-}\right) \equiv -(V_+ + V_-), \quad (13)$$

where $V_+$ and $V_-$ denote the potential energies of the large polarons of radius $R_+$ and $R_-$.[6] The net binding energy of the two oppositely charged large polarons includes positive contributions from the kinetic energies of the self-trapped electron and hole. For hydrogenic self-trapped electronic states the net binding energy for two independent large polarons is $E_{b+} + E_{b-} = (V_+$ and $V_-)/2$ (c.f. page 75 of Ref. 6).

Equation (11) yields $f(s) = 1 + 2/s^2$ for $s \gg 1$. This $s$-dependence of $f(s)$ indicates a potential trough for oppositely charged large-polarons' merger. Thus, as oppositely charged large polarons initially move toward one another the polaron contribution to their potential energy generates an attraction.

A Frenkel exciton starts to form as the oppositely charged large polarons begin to overlap, $S < R_+ + R_-$. In the limit that $S = 0$, the contribution of the net polaron energy from the long-range electron-phonon interaction vanishes since there is then no net charge with which to drive displacements of the surrounding ions. Indeed, Eq. (9) shows that $E^2(S;r,z)$ vanishes for $S =$

0. The exciton's self-trapping then arises from its short-range electron-phonon interaction, the dependence of the exciton's electronic energy on the local expansion or contraction of the solid it contacts.

To address the diminishing long-range electron-phonon interaction associated with exciton formation consider the electric field generated by two oppositely charged points as their separation shrinks. In this regime the components of the electric field in the radial and z-directions generated by the oppositely charged points respectively become:

$$E_r(S \to 0; r, z)/e \approx S \frac{3zr}{(r^2 + z^2)^{5/2}} \quad (14)$$

and

$$E_z(S \to 0; r, z)/e \approx S \frac{(2z^2 - r^2)}{(r^2 + z^2)^{5/2}}. \quad (15)$$

The square of the magnitude of the net carrier-generated electric field in this regime is

$$E^2(S \to 0; r, z)/e^2 \approx S^2 \frac{(r^2 + 4z^2)}{(r^2 + z^2)^4} = S^2 \frac{[1 + 3\cos^2(\theta)]}{r_s^6}, \quad (16)$$

where the cylindrical coordinates are converted to spherical coordinates in the final equality. Integrating this expression outward from a minimum distance $R_m$ associated with overlapping large polarons yields $8\pi s^2/R_m^3$:

$$\lim_{s \to 0} V_{ad}(R_m) = -e^2 \left(\frac{1}{\varepsilon_\infty} - \frac{1}{\varepsilon_0}\right) \frac{S^2}{R_m^3}. \quad (17)$$

Thus, the self-trapping potential associated with the long-range electron-phonon interaction becomes progressively less attractive as the oppositely charged large polarons overlap with one another. In other words, the long-range electron-phonon interactions of electronic carriers with the electric dipoles of the ionic medium that surrounds them then produces a *repulsive* interaction between *oppositely* charged large polarons.

3. Discussion and summary

As schematically illustrated in Fig. 2, to effectively prevent recombination this repulsive interaction must overwhelm the Coulomb attraction between oppositely charged large polarons. To understand how this can occur, recall that large polarons move much more slowly than do conventional quasi-free electronic carriers. This slow motion occurs because large-polaron motion is contingent on appropriate movements of the ions responsible for the self-trapping of their electronic charge carriers. Since large-polaron motion is never faster than ionic motion, the Coulomb attraction between large polarons is screened by their materials' full (static) dielectric constants.[6,12,13] With static dielectric constants being extremely large (e.g. $\varepsilon_0 \approx 1000$ in MAPbX$_3$),[4] the repulsion between oppositely charged large polarons can readily overwhelm their direct Coulomb attraction. Then the net repulsion between oppositely charged large polarons will impede their recombination.

Optically generated pairs of electronic charge carriers do not immediately form large polarons. Time is required for these electronic carriers to become self-trapped in the potential wells established by displacements of the material's ions. In particular, the equilibrium positions of the solid's ions must displace in response to the presence of the electronic charge carriers. The vibrations of these displaced ions must then subsequently relax. Without steric hindrance and bottlenecked relaxation this process can occur in as little as picoseconds.

In summary, the long-range electron-phonon interaction generates displacements of ions' equilibrium positions surrounding electronic charge carriers. This long-range electron-phonon interaction enables electronic charge carriers in multi-dimensional ionic media to form large polarons.[5,6] As large polarons are brought toward one another, the patterns of displaced ions that surround them interfere with one another. This interference is constructive in the region between the two oppositely charged large polarons and destructive otherwise. At large enough separations the net effect of interference between ions' displacement patterns is to produce an attraction between oppositely charged large polarons. Similarly, at large separations interference between ionic displacement induced by large polarons of the same charge produces a repulsion even when they find it energetically favorable to merge into singlet large bipolarons.[11] However, these situations change once the self-trapped electronic carriers of the two large-polarons begin to overlap with one another. Then interference between ionic displacement patterns generates a strong repulsion between oppositely charged large polarons. This repulsion between oppositely charged large polarons will overwhelm their mutual Coulomb attraction if the medium's static dielectric constant is large enough [e.g. Ref. (4) reports $\varepsilon_0 \approx 1000$ in $MAPbX_3$]. Then oppositely charged large polarons will experience a net short-range mutual repulsion. This repulsion between oppositely charged large polaron can forestall their recombination.

Sufficiently large ratios of a medium's static to high-frequency dielectric constants also facilitate like-charged large polarons merging to form large bipolarons.[10,11] The Pauli exclusion principle precludes charge carriers amalgamating into even grander large polarons.[13] In analogy with the situation for oppositely charged large-polarons, oppositely charged large bipolarons will repel one another.

Finally, the physics underlying the formation and interaction of oppositely charged small polarons is compared with that for large polarons. In contrast to the situation for large polarons and bipolarons, the formation of small polarons and bipolarons is typically driven by the short-range electron-phonon interaction.[6] With a short-range electron-phonon interaction, akin to the deformation-potential interaction of covalent solids, the energy of an electronic carrier only depends upon shifts of the atoms which its wavefunction contacts. This situation generates an energy barrier for small-polaron formation.[5,6,18] The long-range electron-phonon interaction that predominates with large-polaron formation does not produce such an energy barrier. These energy barriers introduce time delays for injected or radiation-induced electronic charge carriers forming small polarons.[6,18,19] Like the long-range electron-phonon interaction of large-polaron formation, short-range electron-phonon interactions produce short-range repulsive barriers that impede the recombination of oppositely charged small polarons.[18,19] However, the competitions between indirect short-range repulsion and direct Coulomb attractions can be very different for oppositely charged small- and large-polarons. In particular, the static dielectric constants that control the Coulomb interaction between polarons can be very much larger for large polaron in ionic solids, where $\varepsilon_0 \gg \varepsilon_\infty$, than for small-polarons in covalent solids, where $\varepsilon_0 \approx \varepsilon_\infty$. As such, barriers' to recombination can be more effective for oppositely charged large polarons than for small polarons.

## Acknowledgement

The author gratefully acknowledges the email exchange with Dane deQuilettes which motivated the present work. Specifically, he asked whether the barrier to the recombination between oppositely charged small polarons described in Ref. 18 also applies to large polarons.

## References

1. D. Reagor, E. Ahrens, S.-W. Cheong, A. Migliori and Z. Fisk, Phys. Rev. Lett. **62**, 2014 (1989).

2. G. A. Samara, W. F. Hammetter and E. L. Venturini, Phys. Rev. B **41**, 8974 (1990).

3. G. Cao, J. E. O'Reilly, J. E. Crow and L. R. Testardi, Phys. Rev. B **47**, 11510 (1993).

4. E. J. Juarez-Periz, R. S. Sanchez, L. Badia, G. Garcia-Belmonte, Y. S. Kang, I. Mora-Sero and J. Bisquert, Phys. Chem. Lett. **5**, 2390 (2014).

5. D. Emin and T. Holstein, Phys. Rev. Lett. **36**, 323 (1976).

6. D. Emin, *Polarons* (Cambridge University Press, Cambridge, 2013).

7. H.-B. Schüttler and T. Holstein, Ann. Phys. (N.Y.) **166**, 93 (1986).

8. D. Emin, Phys. Rev. B **45**, 5525 (1992).

9. D. Emin, Phil. Mag. **95**, 918 (2015).

10. D. Emin, Phys. Rev. Lett. **62**, 1544 (1989).

11. D. Emin and M. S. Hillery, Phys. Rev. B **39**, 6575 (1989).

12. D. Emin, Phys. Rev. Lett. **72**, 1052 (1994).

13. D. Emin, Phys. Rev. B. **49**, 9157 (1994).

14. D. Emin, Phil. Mag. **31**, 2931 (2017).

15. S. D. Stranks and H. J. Snaith, Nat. Nanotechnol. **10**, 391 (2015).

16. X.–Y. Zhu and V. Podzorov, J. of Phys. Chem. Lett. **6**, 4758 (2015).

17. K. Miyata, T. L. Atallah and X.–Y. Zhu, Sci. Adv. **3** 1701469 (2017).

18. D. Emin, Physics Today **35** (June), 34 (1982).

19. D. Emin and A. M. Kriman, Phys. Rev. B **34**, 7278, (1986).

Figure Captions

Fig. 1. A positively charged large polaron of radius $R_+$ is separated from a negatively charged large polaron of radius $R_-$ by the distance $S$.

Fig. 2. The general features of the net polaron energy (solid line) and the Coulomb attraction energy (dotted line) for a pair of oppositely charged large polarons is plotted versus $s$, their separation in units of a large-polaron radius $R$. The polaron energy is expressed in units of the polaron energy of independent large polarons. The Coulomb energy is expressed in units of $e^2/\varepsilon_0 R$, where $\varepsilon_0$ represents the static dielectric constant. For sufficiently large static dielectric constants, oppositely charged large polarons will repel one another at short separations thereby impeding their recombination.

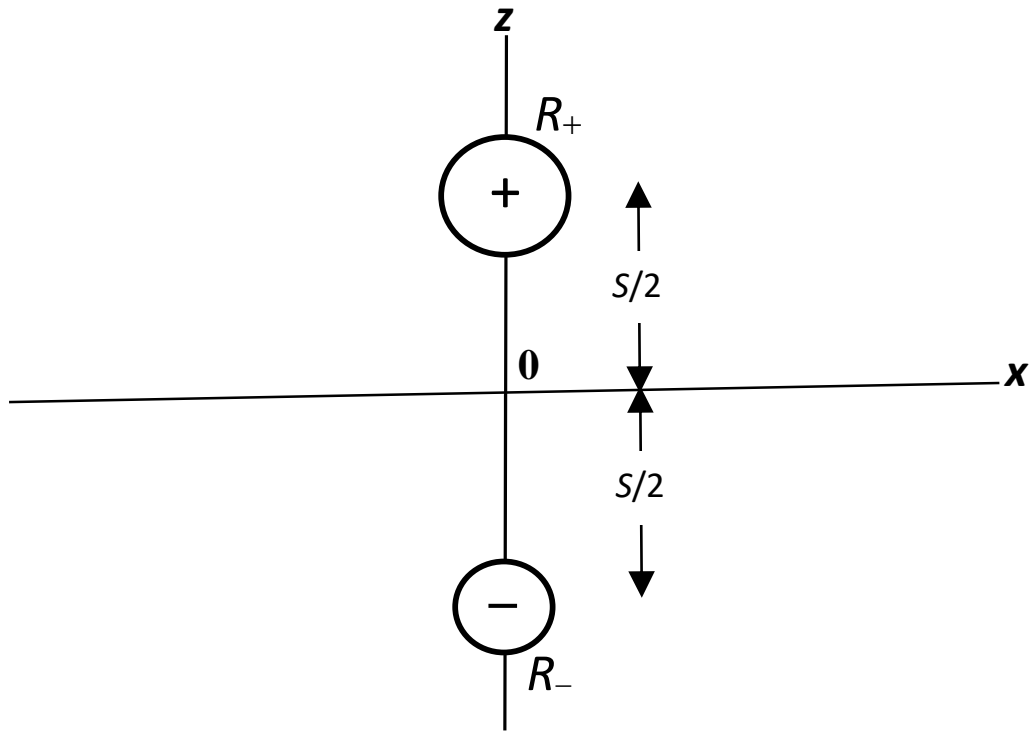

Figure 1

Fig. 1. A positively charged large polaron of radius $R_+$ is separated from a negatively charged large polaron of radius $R_-$ by the distance $S$.

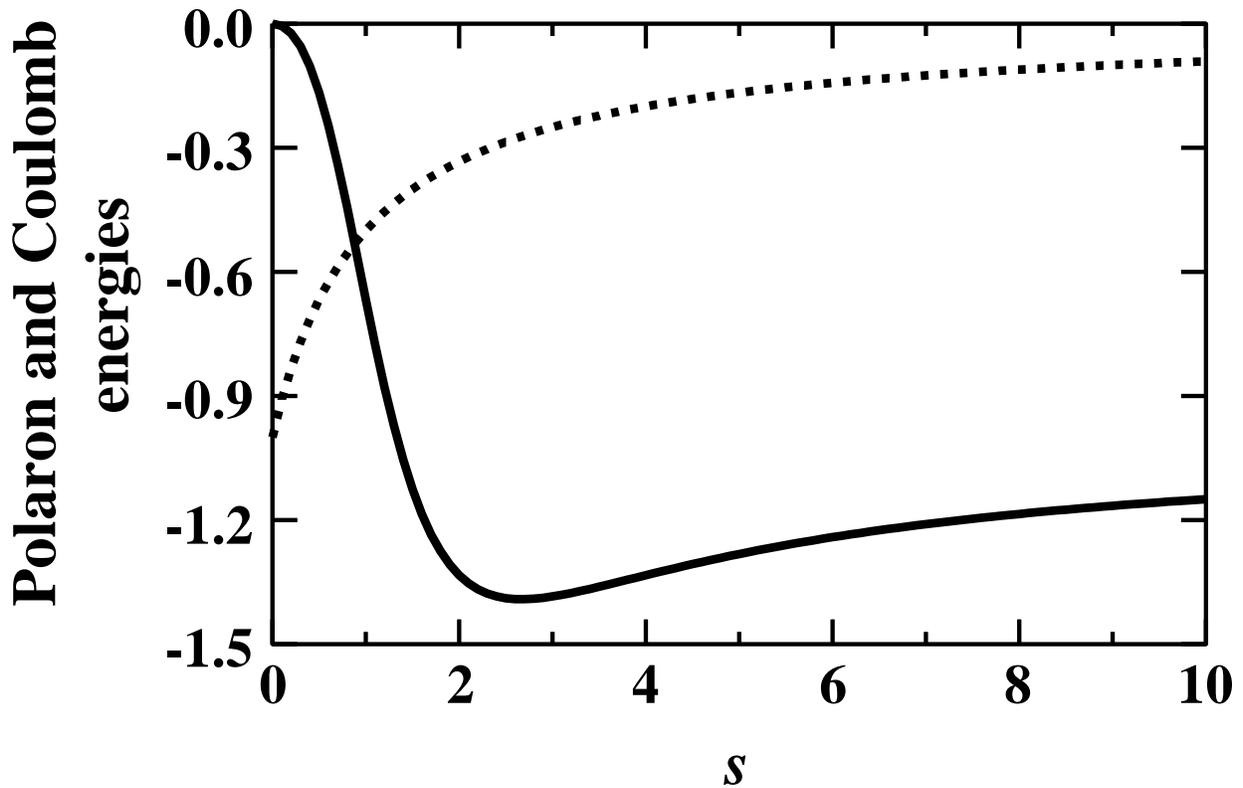

Figure 2

Fig. 2. The general features of the net polaron energy (solid line) and the Coulomb attraction energy (dotted line) for a pair of oppositely charged large polarons is plotted versus $s$, their separation in units of a large-polaron radius $R$. The polaron energy is expressed in units of the polaron energy of independent large polarons. The Coulomb energy is expressed in units of $e^2/\varepsilon_0 R$, where $\varepsilon_0$ represents the static dielectric constant. For sufficiently large static dielectric constants, oppositely charged large polarons will repel one another at short separations thereby impeding their recombination.